\documentclass[11pt]{article}
\usepackage{geometry}
\usepackage{amsfonts}
\usepackage{amsmath}
\usepackage{amssymb}
\usepackage{amsthm}
\usepackage{xcolor}
\usepackage{url}
\usepackage[utf8]{inputenc}
\usepackage{graphicx}
\usepackage{verbatim}
\usepackage{caption,subcaption}
\usepackage{algorithm} %format of the algorithm 
\usepackage{algorithmic} %format of the algorithm 
\usepackage{multirow} %multirow for format of table 

\begin{document}

\title{Deep Predictive Learning of Carotid Stenosis Severity}
\author{Yiqun Diao, Oliver Zhao, Priya Kothapalli, Peter Monteleone, Chandrajit Bajaj}
\date{}

\maketitle
\abstract{
Carotid artery stenosis is the narrowing of carotid arteries, which supplies blood to the neck and head. In this work, we train a model to predict the severity of the stenosis blockage based on SRUC criteria variables and other patient information. We implement classic machine learning methods, decision trees and random forests, used in a previous experiment. In addition, we improve the accuracy through the use of the state-of-art Augmented Neural ODE deep learning method. Through systematic and theory-rooted analysis, we examine different parameters to achieve an accuracy of about 77\%. These results show the strong potential in applying recently developing deep learning methods, while simultaneously suggesting that the current data provided by the SRUC criteria may be insufficient to predict stenosis severity at a high performance level. 
}

\section{Introduction}
Carotid artery stenosis is the narrowing or constriction of carotid arteries, often due to atherosclerosis. While not as accurate as angiography, ultrasound imaging is popular due to its low costs and lack of radiation \cite{Zwiebel1992Duplex} \cite{Dawson1993Role}. Carotid duplex ultrasound is often performed as a screening test on patients who have no signs of carotid artery disease, with the intent to catch asymptomatic carotid artery stenosis at the earliest stage possible \cite{Carroll1991CarotidSonography}\cite{NASCET}. Here, we compare classical machine learning methods such as classification and and regression trees (CART) and random forest with more modern deep neural network methods on the prediction of angiography-measured carotid stenosis severity based on ultrasound-derived data, patient history, and patient demographics. Previously reported methods using decision trees and random forests resulted in a test accuracy ranging from 68-74\%, depending on variables that the model trained on \cite{Peter}. It is shown that the use of novel augmented neural ODEs improves the predictive ability of the carotid data set, compared to the traditional CART and random forest models. 

\section{Background}
\subsection{Carotid Artery Stenosis}
There is a left and right common carotid artery (CCA) that runs along the neck, before branching into the internal carotid arteries (ICA) and external carotid arteries, supplying the brain and face, respectively. The stenosis often occurs distally to the CCA bifurcation due to the local fluid dynamics in the region \cite{Filardi2013Carotid}\cite{Li2019Hemodynamic}. With the internal carotid artery (ICA) supplying the brain, the restriction of blood flow increases the risk of stroke and transient ischemic attack (TIA). A 2010 meta-analysis of 23,000 patients reported ``severe" ($>70\%$) stenosis with a prevalence rate of 1.7\% and 0.9\% in men and women aged 80 or above, respectively \cite{MdeWeerd2010Prevalence}. The Asymptomatic Carotid Surgery Trial 1 (ACST-1) reported an annual stroke risk of 3\% in patients with asymptomatic ($>60\%$) carotid stenosis \cite{Halliday2010tenYear}. The Asymptomatic Carotid Surgery Trial 2 (ACST-2) is currently ongoing \cite{Bulbulia2017ACST}. Another study examining 216 patients with asymptomatic ICA stenosis of 60-99\% over the span of five years put the annual risk of stroke at 3.2\% \cite{Inzitari2000Cause}.

\begin{figure}[!ht]
  \centering
    \includegraphics[width=5cm]{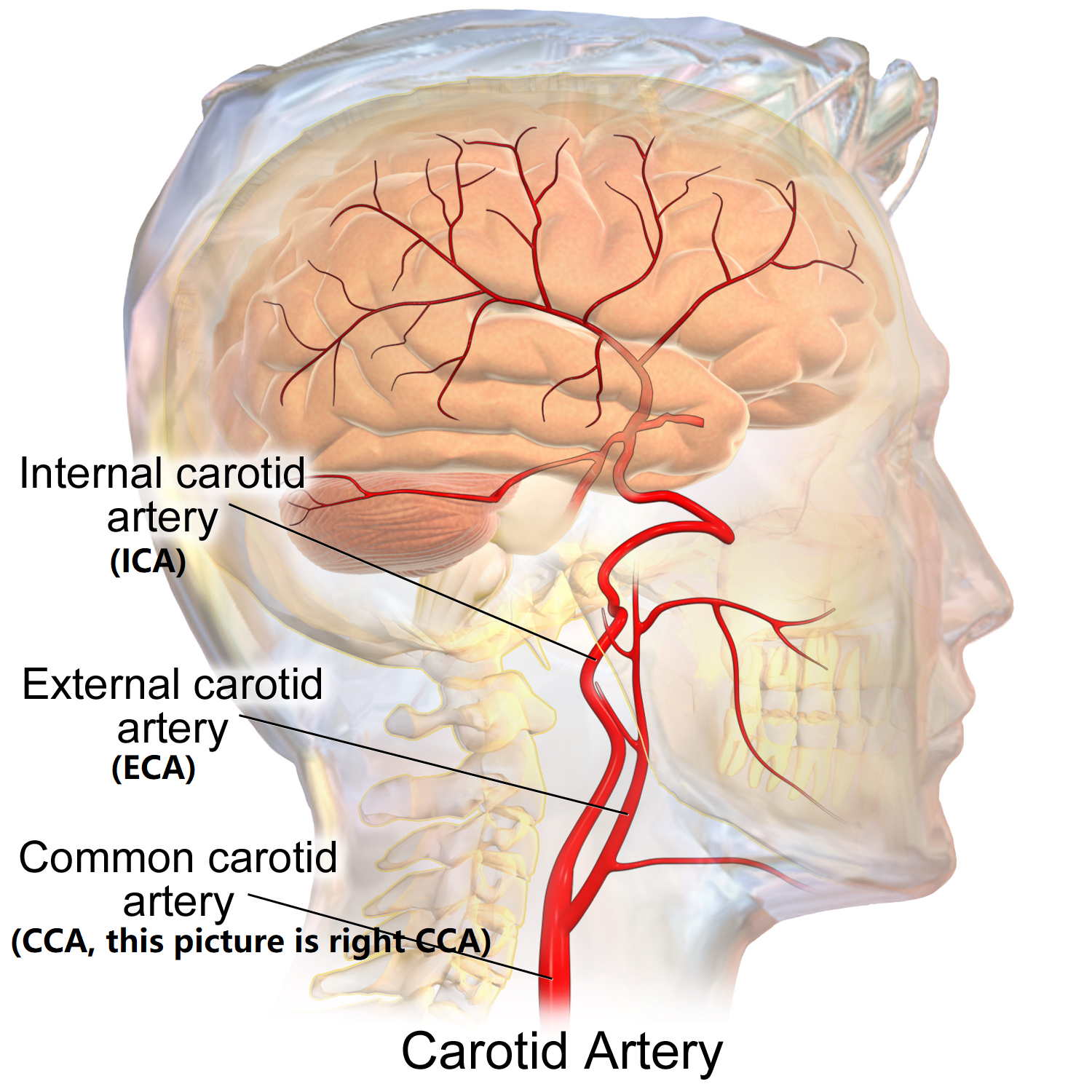}
\caption{Right common carotid artery splitting into the right internal carotid artery and right external carotid artery. } 
\label{1}
\end{figure}

Carotid duplex ultrasound uses Doppler ultrasound to quantitatively assess blood flow velocities and conventional B-mode ultrasound imaging to visualize the artery and accompanying plaque. The severity of carotid artery stenosis is measured through percent blockage of the artery cross-sectional area at the site of plaque. Meanwhile, common blood flow measurements include the three variables used in the Society of Radiologists in Ultrasound Consensus (SRUC) criteria for carotid stenosis: ICA peak systolic velocity (ICA PSV), ICA end diastolic velocity (ICA EDV), and ICA/CCA PSV ratio. \begin{comment}More specifically, the SRUC criteria states that the severity of stenosis is likely to be 70-99\% if the following are true: (1) ICA PSV $\geq$ 230 cm/s, (2) ICA EDV $\geq$ 100, and (3) ICA/CCA PSV Ratio $\geq$ 4.

The PSV is the maximum blood velocity magnitude during the systole and the EDV is the blood velocity at the end of diastole. The PSV and EDV are measured through Doppler ultrasound in centimeters per second (cm/s) through the Doppler effect, where sound waves reflecting off of moving objects induces a frequency shift that is directly correlated with the velocity of the moving objects, as shown below 
\begin{equation}
    F_d = \frac{2F_tv\text{cos}\theta}{c},
\end{equation}
Where $v$ is the velocity of the blood flow, $c$ is the velocity of the sound wave through the tissue, $\theta$ is the angle at which the transducer is applied relative to the direction of blood flow, and $F_d$ and $F_t$ are the detected and transmitted frequencies, respectively. 

\end{comment}
These three SRUC criteria variables are clearly not independent from each other, as these variables are all influenced by blood flow in the internal carotid artery. In our experiments, we will show that while these SRUC variables are valuable in the predictive models, it is suspected that a more comprehensive set of variables are necessary to create a more accurate model.

\subsection{Augmented Neural ODEs (ANODEs)}
Neural networks can approximate some unknown function that relates the inputs to outputs. In our case, the inputs are the patient data and ultrasound variables, and the output is the stenosis severity. A multilayer perceptron, the most basic form of a deep neural network, conveys information through hidden layers in a sequential order. Residual neural networks (ResNets) are a variant of neural networks that utilize skip connections, in which the outputs from previous layers may skip up to several layers to feed directly into the next layer \cite{He2016ResNet}.  

\begin{figure}[!ht]
  \centering
    \includegraphics[width=6cm]{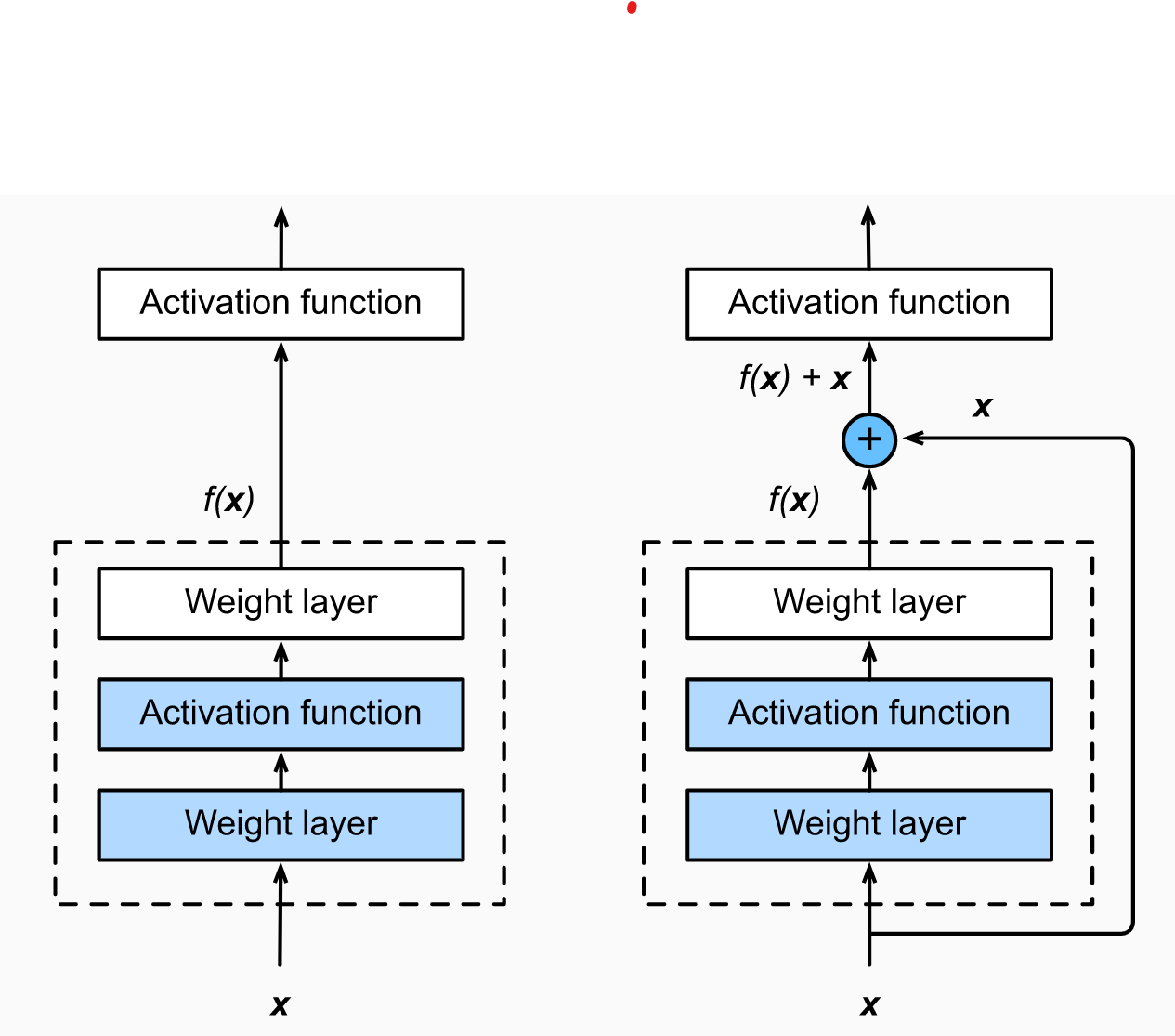}
\caption{On the left we see a regular block used in standard multilayer perceptrons, while on the right we see a residual block. The only difference between the residual block and the regular block is that the residual block allows the inputs $\mathbf{x}$ to skip a hidden layer. Because of the residual block, ResNet can represent more functions than normal neural networks. Figure borrowed from \cite{zhang2019dive}.} 
\label{residual}
\end{figure}

\begin{figure}[!ht]
  \centering
    \includegraphics[width=4cm]{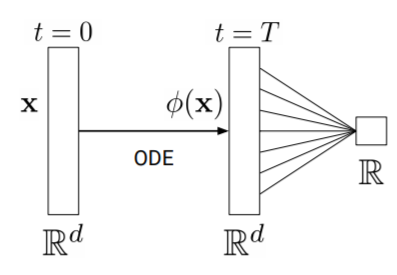}
\caption{Diagram of Neural ODE architecture. Its weights and biases in a Neural ODE are shared across the continuous hidden states, compared to ResNet. So it can represent even more functions \cite{NODEs}.} 
\label{ode-a}
\end{figure}

It has been shown that ResNets effectively act as discretized ordinary differential equations (ODEs), where each residual bock can be written as a forward step Euler discretization \cite{He2016ResNet}. Neural ODE architecture \cite{NODEs} makes it continuous and more powerful. Its diagram is in Figure \ref{ode-a}.
\begin{comment}
For example, a transformation performed by a hidden layer in a ResNet can be mathematically expressed as
\begin{equation}
    \mathbf{h}_{t+1} = \mathbf{h}_t + f(\mathbf{h}_t),
\end{equation}
Where $\mathbf{h}_t \ \epsilon \ \mathbb{R}^d$ is the hidden state in layer $t$ and the hidden layer function $f(\mathbf{h}_t)$ is a differentiable function that maps the inputs from $\mathbb{R}^d$ to $\mathbb{R}^d$. Hence, $\mathbf{h}_{t+1}-\mathbf{h}_t$ acts as a discretized time step for the derivative $\mathbf{h}'_{t}$ with a time step $\Delta t = 1$. Developed in 2018, Neural ODEs (NODEs) expand on ResNet by taking the time step $\Delta t \rightarrow 0$ \cite{NODEs}. In other words, by taking the continuous limit of each discrete layer of the ResNet, the hidden layers become continuous. The continuous hidden states can then be parameterized by an ODE, where 
\begin{equation}
    \underset{\Delta t \rightarrow 0}{\text{lim}} \frac{\mathbf{h}_{t+\Delta t}-\mathbf{h}_t}{\Delta t} = \frac{d\mathbf{h}(t)}{dt} = \mathbf{f}(\mathbf{h}(t),t)
\end{equation}

The NODE can then solve the initial value problem below, mapping a data point $\mathbf{x}$ to a set of features $\phi(\mathbf{x})$ by using an ODE. 
\begin{equation}
    \frac{d\mathbf{h}(t)}{dt} = \mathbf{f}(\mathbf{h}(t),t), \quad \quad \quad \mathbf{h}(0) = \mathbf{x}
\end{equation}
\end{comment}

%The NODE then solves the ODE and learns the corresponding mapped features $\mathbf{h}(T)$. This feature map is then mapped linearly to the output. 
The main advantage of parameterizing hidden states is that unlike ResNet where the weights and biases are time-dependent, the weights and biases in a Neural ODE are shared across the continuous hidden states. Consequently, Neural ODEs can perform with comparable accurracy with ResNet, while using significantly less parameters. Despite the advantages of Neural ODEs, there are several known types of functions that Neural ODEs are not able to train unless at great computational cost, if trainable at all. In acknowledgement of these issues, augmented NODEs (ANODEs) were developed in 2019, where the data is lifted into a higher dimension by a multilayer perceptron before entering the NODE \cite{ANODEs}. The central argument is that by first lifting the data that was initially hard to separate by the NODE, the augmented data is more separable and can be more effectively learned by the NODE.

\section{Methods}
\subsection{Data Set}
The data set contains SRUC criteria, patient demographics, and patient history variables from 257 patients. Patient demographic and patient history variables were excluded with there was not at least 20 samples in each class in a given variable. This resulted in 3 SRUC criteria variables and 23 patient demogaphic and patient history variables. The carotid artery stenosis severity is split into two classes, where a stenosis of $70-99\%$ is a positive case (+1) and a stenosis of $50-69\%$ is a negative case (-1). 

\subsection{Decision Trees and Random Forests}
Machine learning methods for classification such as CART and Random Forest were trained to analyze a clinical carotid data set \cite{Peter}. A decision tree is a flow-like chart in which data samples are separated sequentially based on the threshold of a particular variable. Meanwhile, random forests are a collection of decision trees in which the prediction of each individual tree is used to create a final overall prediction, reducing the bias of individual trees. Because our data set is slightly different from \cite{Peter}, we use the methods applied in the aforementioned study to serve as our own baseline reference. Two decision trees were generated under the following methods: (1) a decision tree that only the SRUC criteria variables and (2) a decision tree that uses patient demographic and patient history variables in addition to the SRUC criteria variables. 

The decision tree splits were determined by measuring the weighted Gini index to maximize class discrimination. The decision tree depth was manually selected to reflect the same depth as the decision trees used previously \cite{Peter}. The reported test accuracy is the average validation accuracy performed across five-fold cross validation. Then, the model was trained on the entire data set to generate the visualized decision trees. Next, we generated two random forest models through similar methods as in the past \cite{Peter}. The first random forest model was trained on only the three SRUC criteria variables, while the second random forest model was trained on all available variables. The number of trees for both models were 100, while the maximum depth of each tree was 2 and 3 for the first and second model, respectively. As before, five-fold cross-validation was performed to provide the training and test accuracy.

\subsection{Deep Learning Method}
In our deep learning experiments, we use the three primary variables from the SRUC criteria, in addition to 23 other variables pertaining to patient demographics and patient history. Patient demographic and patient history variables were excluded if there was not at least 20 samples in each class of a given variable. The data set contains information from 257 patients, in which 200 patient samples were used for the training set and 57 patient samples were used for the testing set.

\begin{figure}[!ht]
  \centering
    \includegraphics[width=4cm]{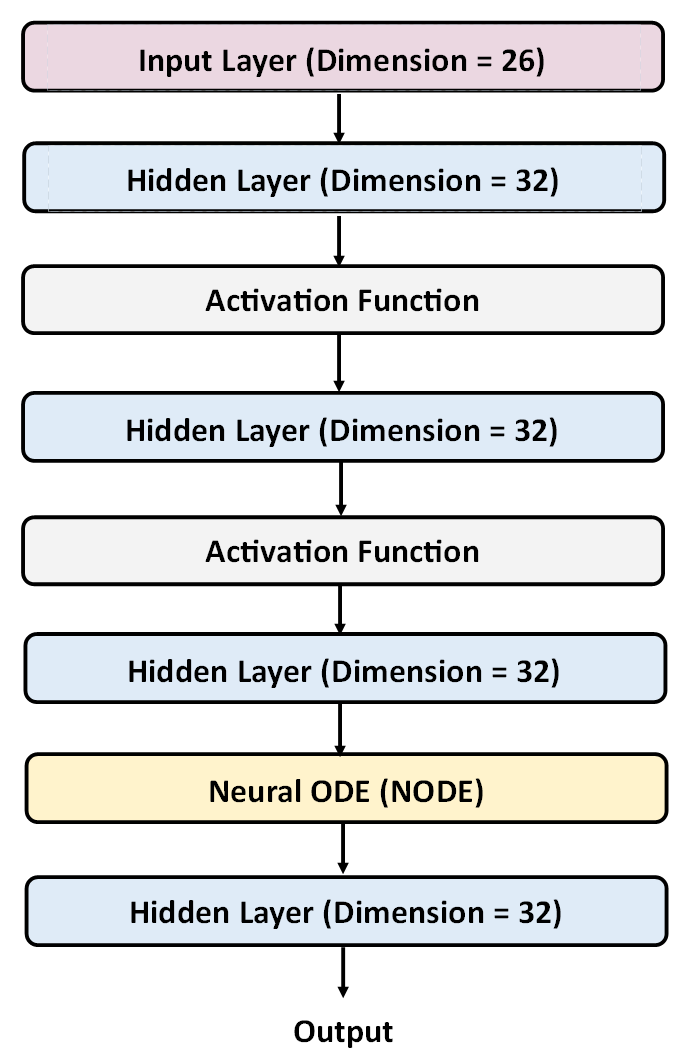}
\caption{General structure of our ANODE adapted to the carotid dataset. ANODE with activation layers have much more functional space than MLP, so it is much more powerful.} 
\label{ANODE}
\end{figure}

In our ANODE, we implement a smooth L1 loss function, which is a piece-wise combination of the L1 and L2 loss functions \cite{Huber}. %, as shown in Equation (\ref{eqn2}). When the absolute value of the argument is close to zero, the Smooth L1 loss function behaves as a L2 loss function. Meanwhile, when the absolute value of the argument is arbitrarily large, the Smooth L1 loss function behaves as a L1 loss function. 
The reason for this distinction is because L1 loss has more steady gradients for large arguments, whereas L2 loss oscillates less for small arguments. %Consequently, the Smooth L1 loss function is mathematically as

%\begin{equation}
%\text{Smooth $L_1$} = 
%\begin{cases} 
%      0.5x^2, & |x|\leq 1 \\
%      |x|-0.5, & |x| > 1
%\end{cases}
%\label{eqn2}
%\end{equation}

We implemented two augmented structures that use either exclusively the ReLU and sigmoid activation functions to map the data from its initial dimension of 26 to a higher dimension of 32 in all hidden layers preceding the NODE. The NODE then projects its output into a 32-dimension vector, whose output provides the prediction as shown in \ref{ANODE}. In the ANODEs, we set $T=1$ with the initial time $t=0$, the batch size to 64, and learning rate to $10^{-3}$. The model is trained for 100 epochs and evaluated with five-fold cross validation. 

\subsection{Visualization of Augmented Data}
We plot the lifted mapping of the three SRUC variables and use principal component analysis (PCA) to represent the high-dimensional, lifted features in a three-dimensional scatter plot. This allows for direct visualization of how the data points are distributed by class. As shown below, by lifting the SRUC variables into higher dimensions, the lifted features are better separated by class. We only show the first batch of the data set, containing 64 instances. The number of epochs was set to 100, with the data displayed at the completion of the last epoch.

\subsection{Canonical Correlation Analysis}
We use canonical-correlation analysis \cite{cano} on the SRUC criteria variables and carotid artery stenosis percentage. Canonical-correlation analysis is a method for establishing relationships or a mapping between two multivariate sets of variables of the same function or domain \cite{CCA}. It could tell us whether the features have a strong correlation with label. 

\subsection{Bias and Variance Analysis in Model Selection}
From \cite{bias}, we know that Min Square Error (MSE) = Bias + Variance + Irreducible Error.
Generally, big bias means underfitting, while high variance means overfitting. If we could make a good balance between bias and variance, we will find an optimal model.

%It is quite often the case that techniques employed to reduce Variance results in an increase in Bias, and vice versa. This phenomenon is called the Bias Variance Trade-off. Balancing the two evils (Bias and Variance) in an optimal way is at the heart of successful model development. 

For Neural Architecture Search, the simplest idea is to use one Neural Architecture Structure to search its best solution. We starts with a simple MLP with only one layer, or a linear classifier. Then each time we add one layer, and of course, test this structure. We continue this until it is very deep (and becomes overfitting), then we choose the best number of layers which has the best validation error on our data set. 

A more complicated idea is to use the binary search method. First, we have a simple model, $S$, as a lower starting point, for example, a linear classifier. For most cases, it would have high bias and underfit. Then, we have a complicated model, $C$, as a higher starting point, for example, a very deep neural network, which would have high variance and overfit. The general idea is shown in Algorithm \ref{alg1}.

\begin{algorithm}[h!]
\caption{NAS with Biase-Variance Trade-off using binary search}  
\label{alg1}
\begin{algorithmic}[1] 
\REQUIRE Simple model $S$, Complicated model $C$
\ENSURE A somewhat better model $B$

\REPEAT 
\STATE M = Find\_Binary\_Structure(S,C)
\STATE Train M on given data set with validation
\IF {M overfits}
\STATE C = M
\ELSIF {M underfits}
\STATE S = M
\ENDIF
\UNTIL{M's test accuracy is satisfactory, for example, several loops of M's test accuracy almost converges, or their test accuracy does not vary much} 
\STATE Output a better model B = M

\end{algorithmic} 
\end{algorithm}

As to the metric judging overfitting and underfitting, we could use generalization error, i.e., train accuracy - validation accuracy. If it exceeds some threshold, we assume it as overfitting. In our experiment, we set the threshold as 3\%.

\subsection{Energy-based learning and search of loss function}
The core idea of energy-based learning \cite{2006A} is that the final network is an energy function. For an input data, we find a label (classification) or prediction value (regression) that minimizes the energy. The loss function is controlling the adaptation of energy function, or measuring the quality of energy function. So the choice of loss function is the criterion to measure the energy function. 

From the conclusion in the perspective of energy learning \cite{2006A}, log-likelihood loss function is better. It is due to the shape of energy function. Log-likelihood loss function is more concave and could converge more quickly. It is proved that generalized margin loss is also a relatively better loss function. The difference is that log-likelihood loss considers all possible labels, while margin loss only considers the correct and most offending incorrect answer. By setting a margin between correct answer and most offending incorrect answer, one can guarantee that the shape of energy function would not be too flat, which avoids a tie of different labels in the prediction. \cite{enreg} also shows that entropy regularization can help find a better model with fast convergence. 

For carotid data set, since we only have 2 classes, margin loss is enough. The search space enlarges because of the trade-off parameter between cross entropy loss and margin loss.

\section{Results}
\subsection{Decision Trees and Random Forests}
When training the decision tree on the three primary SRUC variables with the freedom to select any threshold, the training accuracy was 76.6526 $\pm$ 0.9954\% and the testing accuracy was 73.1373 $\pm$ 5.2375\%. When training the decision tree on all available variables - ultrasound variables, patient demographics, and patient history - the training accuracy was 79.7656 $\pm$ 0.9211\% and the testing accuracy was 72.3680 $\pm$ 4.0209\%. Meanwhile, the decision tree trained on all available data contains a slightly lower accuracy as a result of not predicting every single case to belong to the same class of 70-99\% stenosis. In the random forest model trained only on the SRUC variables, the model had a training accuracy of 77.0438 $\pm$ 0.8066 and testing accuracy of 74.7134 $\pm$ 2.7088. In the random forest model trained on all available variables, the model had a training accuracy of 77.7239 $\pm$ 0.6366 and testing accuracy of 75.0974 $\pm$ 0.9421.

\begin{figure}[!ht]
  \centering
    \includegraphics[width=7cm]{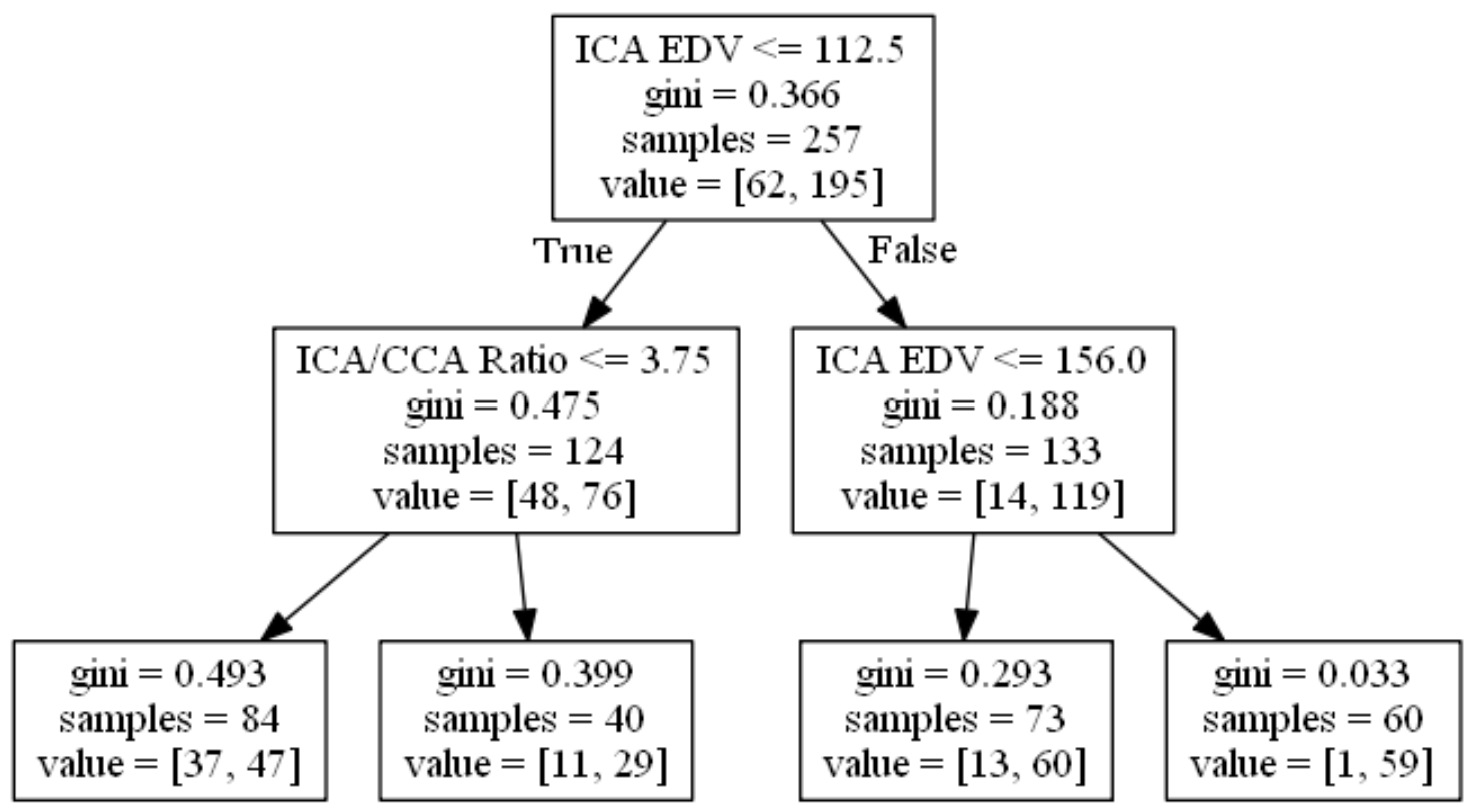}
\caption{Decision tree trained on SRUC variables only.} 
\label{1}
\end{figure}

\begin{figure}[!ht]
  \centering
    \includegraphics[width=13cm]{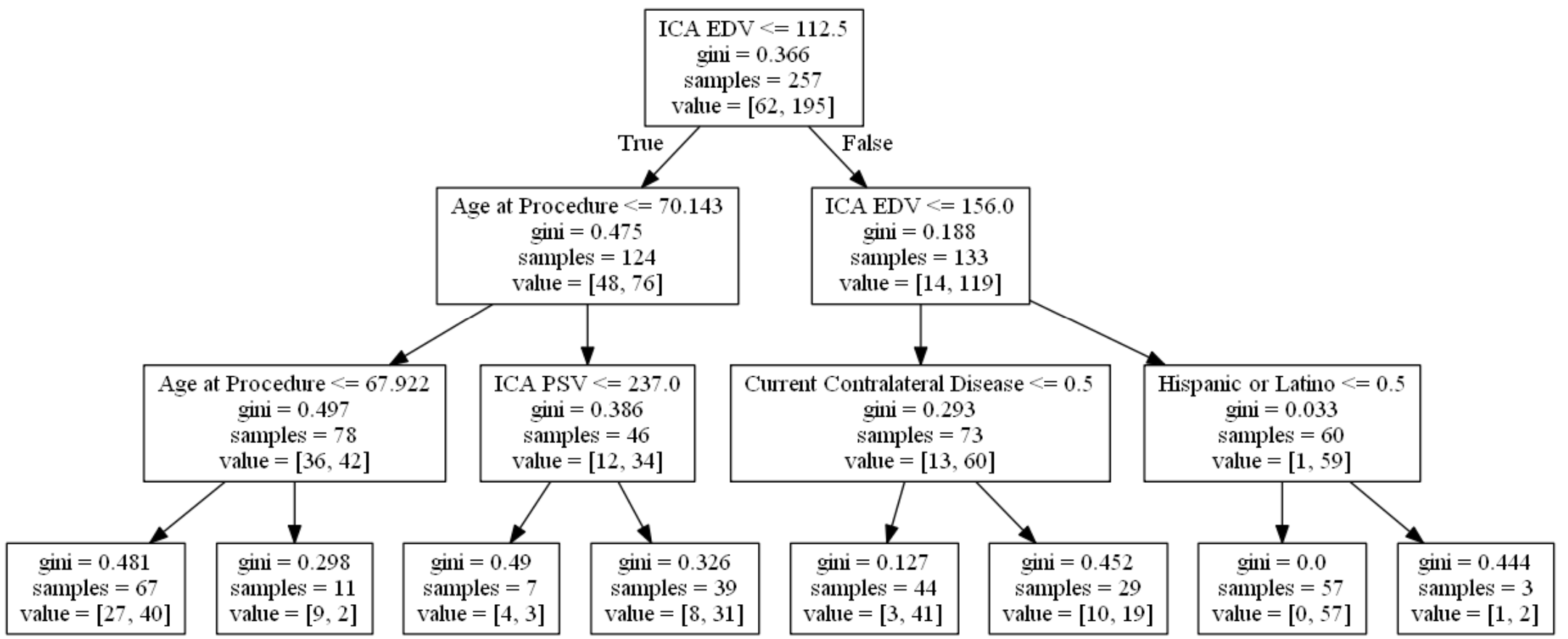}
\caption{Decision tree trained on all available variables in the data set: SRUC variables, patient demographics, and patient history.} 
\label{2}
\end{figure}

\subsection{Deep Learning with ANODEs}

When using both the ReLU and sigmoid activation functions, the test accuracy obtained was approximately 77\%, slightly better than decision tree method. The ReLU activation function showed significant overfitting, while the sigmoid activation function did not. The ReLU activation function results in a training error of nearly zero and a testing error of about 22\%. The false negative rate in the test set is about 18\%. Meanwhile, the sigmoid activation function provides better generalization, with both the training and testing error at about 23\%. The false negative rate in the testing set for the sigmoid activation function is nearly zero. These results suggest that the sigmoid activation function is a better regularizar than the ReLU activation function. 

\begin{figure}[!ht]
  \centering
    \includegraphics[width=12cm]{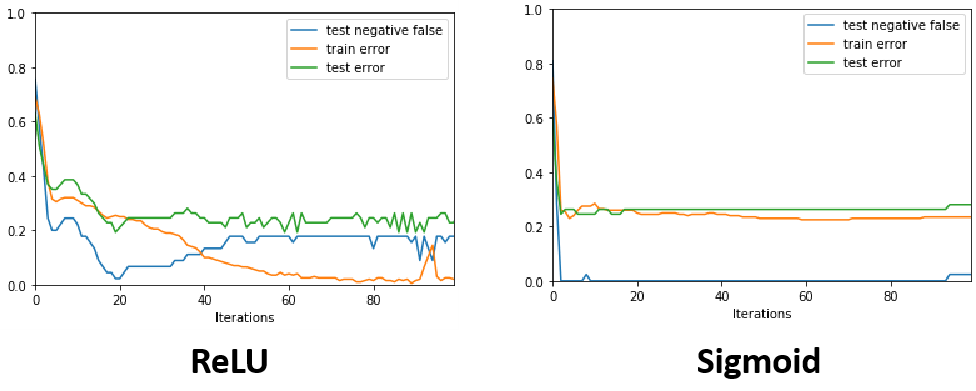}
\caption{Training error, testing error, and false negative rate in the test sets with ReLU and sigmoid activation functions. We can conclude that (1) ReLU is better than Sigmoid; (2) deep learning method is slightly better than decision tree.} 
\label{26acc}
\end{figure}

The best separation achieved when the data is augmented to an additional 6 dimensions. When augmenting the data to an addition 4 dimensions, we see improved separation between the two classes. However, there is still significant overlap. Meanwhile, we see that augmenting the data to an additional 6 dimensions results in better separation. Lastly, augmenting the data to an additional 9 dimensions shows poor separation, suggesting that overfitting has occurred. 

\begin{figure}[!ht]
  \centering
    \includegraphics[width=12cm]{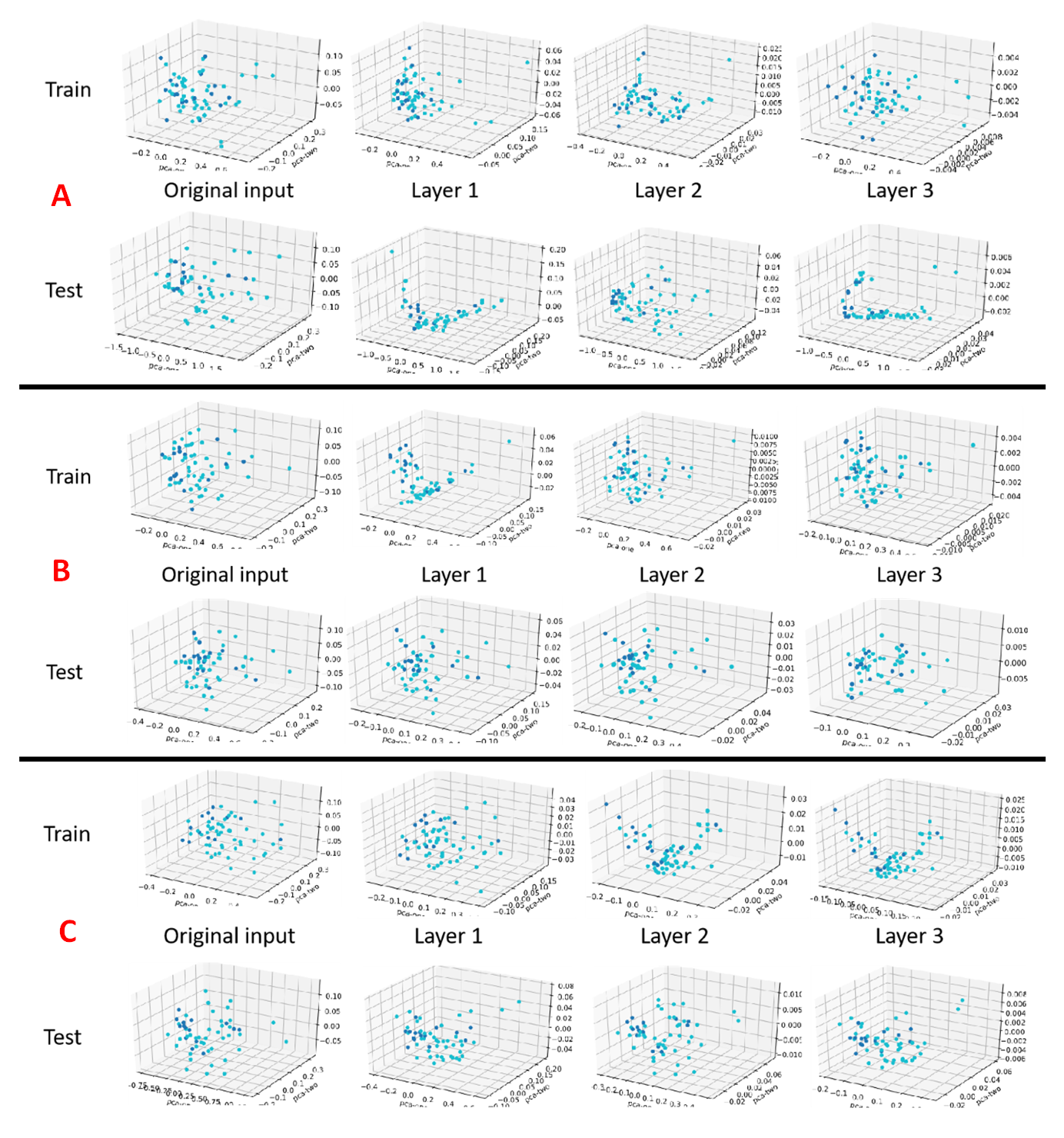}
\caption{Transformation of data points when augmented with an additional: (A) 4 dimensions, (B) 6 dimensions, and (C) 9 dimensions prior to entering the NODE. Dark blue indicates a positive class with a stenosis of 70-99\%, while light blue indicates a negative class with a stenosis of 50-69\%. ReLU activation function used. It shows that the augmented layer cannot separate those two parties well, even with powerful Neural ODE Net.} 
\label{aug4}
\end{figure}

\begin{comment}

When augmenting the data to an additional 6 dimensions with a ReLU activation function, we visualize the transformation of the data points in each layer in Figure \ref{disi} and examine how the gradient affects the parameters in each layer in Figure \ref{grad}.

\begin{figure}[!ht]
  \centering
    \includegraphics[width=12cm]{dis.png}
\caption{This picture shows the distance changes through each layer when augmented to 6 dimensions. For each layer, it calculates 3 values. The first value is the average distance from point of Class 0 to its center, the second value is the distance from point of Class 1 to its center, the third is the distance between centers of 2 classes. The picture shows that they are all decreasing, but the distance between the two classes is smaller than other 2 values, which means that the transformation is separating the 2 classes.} 
\label{disi}
\end{figure}

\end{comment}

\begin{figure}[!ht]
  \centering
    \includegraphics[width=12cm]{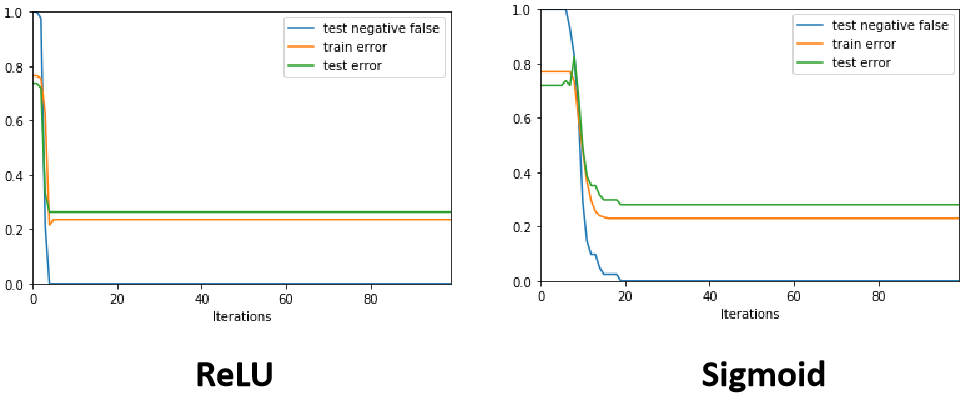}
\caption{This picture shows the training process only using SRUC criterion. As is shown, both ReLU and Sigmoid activation function lead to a flat line, where the test error is about 25\%. Because the amount of data are so limited, it can hardly find a connection to divide. And it has zero negative false and tends to predict as severe stenosis block.} 
\label{3acc}
\end{figure}

\begin{figure}[!ht]
  \centering
    \includegraphics[width=6cm]{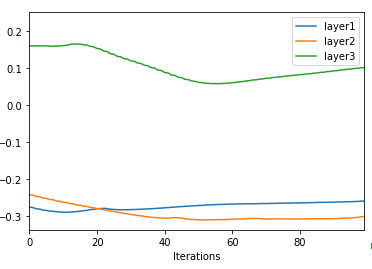}
\caption{Gradient changes for a parameter in each layer. It is augmented from 3 inputs to 6 dimensions using ReLU activation. As is indicated in the picture, the lifted mapping is working because the gradient is changing steadily. However it also shows the gradient vanishing, which means the variation of parameters become smaller as the layer goes down. For our example, however, the gradient still works, which indicates that the depth is proper. If we make the network deeper, then the gradient will vanish to almost 0.} 
\label{grad}
\end{figure}

For the choice of activation function, ReLU has a tendency for better fitting. Prevalent views also regard ReLU as better activation function because it will tackle the gradient problem. However in this dataset, ReLU results in an overfitted neural network. Also when combined with batch norm, the sigmoid activation function performs better than ReLU. %This is summarized by our experiments with the multilayer perceptron before using the Augmented Neural OdeNet. After batch normalization, about half of the data features become negative. This is problematic for the ReLU activation function, as it results in dead nodes in which the output becomes zero. In contrast, the sigmoid function provides more smooth mapping and avoids the issue of the ``dying ReLU". Additionally, with this specific data set the sigmoid function appears to work well as a regularizer and prevents the training accuracy from rising too accurately, helping improve the generalization of the model. 

%With regards to loss functions, the L1 Smooth loos function performed well. This is in contrast to the hinge loss function used typically for support vector machines (SVMs), which performed poorly. When we choose the default L1SmoothLoss, it performs well. But the hinge loss for SVM isn’t working for this scenario. 

%When utilizing a traditional multilayer perceptron, the training accuracy fluctuates significantly throughout successive epochs. Even after adding regularization method such as batch normalization, weight decay, and dropout, this issue remained. This is most likely due to the entangled data points within the feature space, resulting in a complex and unstable neural network. Now with Augmented Neural OdeNet, the data can be projected into a higher dimension with the data points in a more separable state, allowing for more effective training and consequently better results. 

Our work with Augmented Neural ODEs provides better performance than previous classical machine learning models. Although the current results are not satisfactory for diagnosis in clinical settings, it is shown that the use of more modern machine learning methods can provide modest improvements. In light of the complexity and robustness of the Augmented Neural ODEs, these results suggest that the current SRUC criteria variables may be insufficient to accurately predict the severity of carotid artery stenosis. By incorporating more thorough ultrasound-derived data, it is likely that the ability to discriminate between different levels of carotid stenosis severity will significantly increase.

\subsection{Canonical-Correlation Analysis}
The correlation coefficient between the SRUC criteria and carotid stenosis percentage is 0.329. If we combine the SRUC criteria with the available 23 patient history and demographic variables, the correlation coefficient increases to 0.407. It is only weak linear correlation, which further proves that the features in carotid data set could not be a strong indicator for the stenosis artery severity.

\subsection{Bias and Variance Analysis in Model Selection}
For carotid data set, we first perform the idea of linear search for the depth of hidden layers. We automatically add a hidden layer to the model, one by one. Then we train on each model, getting the result of train error and test error. For model choice purpose, we want a balance of bias and variance. Bias means train error, variance means generalization error. When train acc $>$ test acc, bias + variance = test error; when train acc $<$ test acc, bias = train error, but variance cannot be smaller than 0. So we use train error to denote it. We would like to minimize bias + variance, in this analysis, we should maximize min\{train acc, test acc\}. This is the criterion we choose.

Here is the experimental result we get for carotid data set. The best chosen model has 2 layers, where train accuracy = 75.5\%, test accuracy = 77.2\%. The best results are similar to that of Augmented Neural ODE Net. 

Then we also try cross-validation method. We randomly divide the data set into training set and validation set, for several times. Then we use the gap between training accuracy and validation accuracy as the indicator of generalization error.
In our setting, we use three-folder cross validation. Results are shown in Table \ref{table-acc}:

\begin{table}[ht]
   \caption{Cross-validation results of model with different number of layers. All of these model can get only about 77\% test accuracy. }
   \label{table-acc}
   \begin{center}
        \begin{tabular}{| c | c | c |}
        \hline
              Number of Layers & Avg Train Acc & Avg Validation Acc \\
        \hline
              1 & 0.767 & 0.760\\
        \hline
              2 & 0.760 & 0.754\\
        \hline
              3 & 0.753 & 0.778\\
        \hline
              4 & 0.753 & 0.778\\
        \hline
              5 & 0.755 & 0.771\\
        \hline
        \end{tabular}
    \end{center}
\end{table}

Further, we consider both depth (in the step of 1) and number of neurons (in the step of 4). We try depth from 1 to 5, number of neurons from 8 to 32. Since carotid data set is small, we just search all those neural architecture spaces. The criterion for evaluating bias and variance is the same as previous sections. As a result, the program returns the best depth as 2, the best number of neurons as 16, where training accuracy is 0.77 and test accuracy is 0.772. Although we have searched for so large neural architecture space to balance bias and variance, the test accuracy still fails to improve any more. This could serve as another evidence that SRUC variables could not be a good indicator of carotid artery severity. 

\subsection{Search of loss function}
We add the margin loss to the previous cross entropy loss. Besides the depth and number of neurons, we also search the trade-off parameter between cross entropy loss and margin loss. We try the trade-off parameter in  \{0.05, 0.1, 0.2, 0.5\}. The best result is shown in Table \ref{table-loss}.

\begin{table}[ht]
   \caption{Neural architecture searching results of different trade-off parameter for the margin loss. The best test accuracy is just 78.9\%, only a little bit better than 77\%.}
   \label{table-loss}
   \begin{center}
        \begin{tabular}{| c | c | c |}
        \hline
              Trade-off parameter & Best Train Acc & Best Test Acc \\
        \hline
              0.05 & 0.755 & 0.719\\
        \hline
              0.1 & 0.765 & 0.789\\
        \hline
              0.2 & 0.775 & 0.772\\
        \hline
              0.5 & 0.765 & 0.789\\
        \hline
        \end{tabular}
    \end{center}
\end{table}

As we can see, all results are still around 77\% accuracy. It further proves that even with more powerful margin loss, we cannot get better results for carotid data set. From all these experimental results above, we can conclude that current features cannot be a good indicator to the carotid artery severity. 

\section{Discussion}
Unsurprisingly, the random forest outperforms the decision tree, resulting in a testing accuracy of 75.1\% when using SRUC variables in addition to the patient demographic and history variables. When only using the SRUC variables, the random forest testing accuracy drops to 74.7\%. This suggests that the use of patient demographic and history variables may be beneficial in providing more predictive power. The impact of these variables may be more evident if the model is trained on a larger patient population. 

Meanwhile, we see that implementing our final ANODE model results in a testing error of about 77\%, which is a modest improvement over the random forest models. It is believed that this improvement is due to the ANODEs ability to better separate data points through augmented dimensions and the NODEs general ability to capture complex relationships between the features. However, an accuracy of 77\% is insufficient for use in assisted diagnostics. The general ability of ANODEs to capture complex relationships, in contrast to the relatively poor performance, suggests that the current SRUC criteria may be insufficient in satisfactorily determining the severity of carotid artery stenosis.

We use Canonical-Correlation Analysis to explore the possibility that the data set may not contain enough discriminatory information to accurately predict carotide artery stenosis. The weak correlation coefficient of 0.329 between the SRUC variables and carotid stenosis percentages indicates that the current SRUC criteria variables are lacking in discriminatory power to capture the severity of carotid artery stenosis. Even with the addition of patient history and demographics, the correlation coefficient is still weak at 0.407. Also, Neural Architecture Search tries to find an optimal structure of neural network for carotid data set, but none of those structures could exceed 77\% test accuracy. These results support the possibility that the SRUC criteria is insufficient in determining carotid artery stenosis severity. To improve future variants of the ANODE, it is strongly believed that having more information ultrasound-derived information outside of the SRUC criteria will provide an significant increase in discriminatory power.

However, another possibility for relatively low performance of the machine learning models is that the current data set is lacking in size and class balance. Consequently, an important step forward would be increasing the sample size of the data set, in addition to increasing the number of ultrasound-derived variables. 

\section{Conclusion}
This work aims to develop a model to predict the severity of stenosis blockage based on the SRUC criteria variables and other patient information. We implement the previously used classic machine learning methods to serve as a baseline reference \cite{Peter}. We improve the accuracy through ANODEs to achieve a final accuracy of about 77\%. Through our theoretical analyses with Canonical Correlation Analysis and experiments with different regularization methods and neural network architectures, we believe our model performs well when considering the constraints of the data set. We see that the deep learning method is able to capture more complex relationships to provide more predictive power. Despite the improved results, there is still a need for significantly larger data set with more patients and variables before using it for clinical applications.

\emph{Acknowledgement:} 
 %acknowledgements go in this section.
 This research was supported in part by a grant from NIH - R01GM117594, by the Peter O’Donnell Foundation and in part from a grant from the Army Research Office accomplished under Cooperative Agreement Number W911NF-19-2-0333. The work of A.R. is supported by U.S. Department of Energy, Office of High
Energy Physics under Grant No. DE-SC0007890 The views and conclusions contained in this document are those of the authors and should not be interpreted as representing the official policies, either expressed or implied, of the Army Research Office or the U.S. Government. The U.S. Government is authorized to reproduce and distribute reprints for Government purposes notwithstanding any copyright notation herein.

\bibliographystyle{plain}
\bibliography{references}

\begin{thebibliography}{10}

\bibitem{NASCET}
North american symptomatic carotid endarterectomy trial: Methods, patient
  characteristics, and progress.
\newblock {\em Stroke}, 22:711--720, 1991.

\bibitem{bias}
Anand Avati.
\newblock Bias-variance analysis: Theory and practice.

\bibitem{cano}
Magnus Borga.
\newblock Canonical correlation a tutorial.

\bibitem{Bulbulia2017ACST}
Richard Bulbulia and Alison Halliday.
\newblock The asymptomatic carotid surgery trial-2 (acst-2): an ongoing
  randomised controlled trial comparing carotid endarterectomy with carotid
  artery stenting to prevent stroke.
\newblock {\em Health Technology Assessment}, 21:1--40, 2017.

\bibitem{Carroll1991CarotidSonography}
Barbara~A. Carroll.
\newblock Carotid sonography.
\newblock {\em Radiology}, 178:303--313, 1991.

\bibitem{NODEs}
Ricky T.~Q. Chen, Yulia Rubanova, Jesse Bettencourt, and David Duvenaud.
\newblock Neural ordinary differential equations.
\newblock {\em 32nd Conference on Neural Information Processing Systems}, 2018.

\bibitem{CCA}
John Shawe-Taylor David R.~Hardoon, Sandow~Szedmak.
\newblock Canonical correlation analysis: An overview with application to
  learning methods.
\newblock {\em Neural Computation}, pages 2639--2664, 2004.

\bibitem{Dawson1993Role}
David~L. Dawson, R.~Eugene Zierler, Jr. D.~Eugene~Strandness, Alexander~W.
  Clowes, and Ted~R. Kohler.
\newblock The role of duplex scanning and arteriography before carotid
  endarterectomy: A prospective study.
\newblock {\em Journal of Vascular Surgery}, 18:673--683, 1993.

\bibitem{MdeWeerd2010Prevalence}
de~Weerd~M, Greving JP, Hedblad B, Lorenz MW, O'Leary DH, Rosvall M, Sitzer M,
  Buskens E, and Bots ML.
\newblock Prevalence of asymptomatic carotid artery stenosis in the general
  population: an individual participant data meta-analysis.
\newblock {\em Stroke}, 41:1294--1297, 2010.

\bibitem{ANODEs}
Emilien Dupont, Arnaud Doucet, and Yee~Whye Teh.
\newblock Augmented neural odes.
\newblock {\em NIPS Proceedings}, 2019.

\bibitem{Filardi2013Carotid}
Vincenzo Filardi.
\newblock Carotid artery stenosis near a bifurcation investigated by fluid
  dynamic analyses.
\newblock {\em The Neuroradiology Journal}, 26:Epub, 2013.

\bibitem{Halliday2010tenYear}
Alison Halliday, Michael Harrison, Elizabeth Hayter, Xiangling Kong, Averil
  Mansfield, Joanna Marro, Hongchao Pan, Richard Peto, John Potter, Kazem
  Rahimi, Angela Rau, Steven Robertson, Jonathan Streifler, and Dafydd Thomas.
\newblock 10-year stroke prevention after successful carotid endarterectomy for
  asymptomatic stenosis (acst-1): a multicentre ranodmised trial.
\newblock {\em The Lancet}, 376:1074--1084, 2010.

\bibitem{He2016ResNet}
Kaiming He, Xiangyu Zhang, Shaoqing Ren, and Jian Sun.
\newblock Deep residual learning for image recognition.
\newblock {\em Proceedings of the IEEE conference on computer vision and
  pattern recognition}, pages 770--778, 2016.

\bibitem{Huber}
Peter~J. Huber.
\newblock Robust estimation of a location parameter.
\newblock {\em Annals of Statistics}, 53:73--101, 1964.

\bibitem{Inzitari2000Cause}
Domenico Inzitari, Michael Eliasziw, Peter Gates, Brenda~L. Sharpe,
  Richard~K.T. Chan, Heather~E. Meldrum, and Henry~J.M. Barnett.
\newblock The causes and risk of stroke in patients with asymptomatic
  internal-carotid-artery stenosis.
\newblock {\em The New England Journal of Medicine}, 342:1693--1700, 2000.

\bibitem{2006A}
Yann Lecun, Sumit Chopra, Raia Hadsell, Marc'Aurelio Ranzato, and Fu~Jie Huang.
\newblock A tutorial on energy-based learning.
\newblock In {\em Predicting Structured Data}, 2006.

\bibitem{Li2019Hemodynamic}
Cong-Hui Li, Bu-Lang Gao, Ji-Wei Wang, Jian-Feng Liu, Hui Li, and Song-Tao
  Yang.
\newblock Hemodynamic factors affecting carotid sinus atherosclerotic stenosis.
\newblock {\em World Neurosurgery}, Epub, 2019.

\bibitem{Peter}
Peter~P. Monteleone, Ido Weinberg, Brian Allen, Daniel~Roy Miller, Andrew Ward,
  and David Scheinker.
\newblock Machine learning prediction of carotid stenosis severity by vascular
  ultrasound.
\newblock {\em In Review}.

\bibitem{enreg}
Yuxin Chen Yuting Wei Yuejie~Chi Shicong~Cen, Chen~Cheng.
\newblock Fast global convergence of natural policy gradient methods with
  entropy regularization.
\newblock 2020.

\bibitem{zhang2019dive}
Aston Zhang, Zachary~C. Lipton, Mu~Li, and Alexander~J. Smola.
\newblock {\em Dive into Deep Learning}.
\newblock 2019.
\newblock \url{http://www.d2l.ai}.

\bibitem{Zwiebel1992Duplex}
William~J. Zwiebel.
\newblock Duplex sonography of the cerebral arteries: efficacy, limitations,
  and indications.
\newblock {\em American Journal of Roentgenology}, 158:29--36, 1992.

\end{thebibliography}
%\CBnote{Please could either or both convert all references from the earlier paper of Peter Monteleone,  into .bib format . they will then get appropriately cited in this paper.}

\end{document}